\documentclass[12pt]{article}
\usepackage{amssymb}
\renewcommand{\theequation}{\arabic{section}.\arabic{equation}}

\begin{document}

%************************** Text Begins here ******************************

%  Greek letters

\def\a{\alpha}
\def\b{\beta}
\def\d{\delta}
\def\e{\epsilon}
\def\g{\gamma}
\def\h{\mathfrak{h}}
\def\k{\kappa}
\def\l{\lambda}
\def\o{\omega}
\def\p{\wp}
\def\r{\rho}
\def\t{\tau}
\def\s{\sigma}
\def\z{\zeta}
\def\x{\xi}
 \def\A{{\cal{A}}}
 \def\B{{\cal{B}}}
 \def\C{{\cal{C}}}
 \def\D{{\cal{D}}}
\def\G{\Gamma}
\def\K{{\cal{K}}}
\def\O{\Omega}
\def\R{\bar{R}}
\def\T{{\cal{T}}}
\def\L{\Lambda}
\def\f{E_{\tau,\eta}(sl_2)}
\def\E{E_{\tau,\eta}(sl_n)}
\def\Zb{\mathbb{Z}}
\def\Cb{\mathbb{C}}

\def\R{\overline{R}}
% Shorthands for \begin{equation} and the like

\def\beq{\begin{equation}}
\def\eeq{\end{equation}}
\def\bea{\begin{eqnarray}}
\def\eea{\end{eqnarray}}
\def\ba{\begin{array}}
\def\ea{\end{array}}
\def\no{\nonumber}
\def\le{\langle}
\def\re{\rangle}
\def\lt{\left}
\def\rt{\right}

\newtheorem{Theorem}{Theorem}
\newtheorem{Definition}{Definition}
\newtheorem{Proposition}{Proposition}
\newtheorem{Lemma}{Lemma}
\newtheorem{Corollary}{Corollary}
\newcommand{\proof}[1]{{\bf Proof. }
        #1\begin{flushright}$\Box$\end{flushright}}

\baselineskip=20pt

%%%%%%%%%%%%%%%%%%%%%%%%%%%%%%%%%%%%%%%%%%%%%%%%%%%%%%%%%%%%
%                                                          %
%  Title page                                              %
%                                                          %
%%%%%%%%%%%%%%%%%%%%%%%%%%%%%%%%%%%%%%%%%%%%%%%%%%%%%%%%%%%%
\newfont{\elevenmib}{cmmib10 scaled\magstep1}
\newcommand{\preprint}{
   \begin{flushleft}
     \elevenmib Yukawa\, Institute\, Kyoto\\
   \end{flushleft}\vspace{-1.3cm}
   \begin{flushright}\normalsize
   \sf  YITP-05-37\\
     {\tt hep-th/0507148} \\ July 2005
   \end{flushright}}
\newcommand{\Title}[1]{{\baselineskip=26pt
   \begin{center} \Large \bf #1 \\ \ \\ \end{center}}}
\newcommand{\Author}{\begin{center}
   \large \bf
Wen-Li Yang$,{}^{a,b}$
 ~ Yao-Zhong Zhang ${}^b$ and~Ryu Sasaki${\,}^c$\end{center}}
\newcommand{\Address}{\begin{center}

     ${}^a$ Institute of Modern Physics, Northwest University,
     Xian 710069, P.R. China\\
     ${}^b$ Department of Mathematics, University of Queensland, Brisbane, QLD 4072,
     Australia\\
     ${}^c$ Yukawa Institute for Theoretical Physics, Kyoto
     University, Kyoto 606-8502, Japan
   \end{center}}
\newcommand{\Accepted}[1]{\begin{center}
   {\large \sf #1}\\ \vspace{1mm}{\small \sf Accepted for Publication}
   \end{center}}

\preprint
\thispagestyle{empty}
\bigskip\bigskip\bigskip

\Title{$A_{n-1}$  Gaudin
      model with  open boundaries } \Author

\Address
\vspace{1cm}

\begin{abstract}
The $A_{n-1}$ Gaudin model with  integerable boundaries specified
by  {\it non-diagonal\/} K-matrices is studied. The commuting
families of Gaudin operators are diagonalized by the algebraic
Bethe ansatz method. The eigenvalues and the corresponding Bethe
ansatz equations are obtained.

\vspace{1truecm} \noindent {\it PACS:} 03.65.Fd; 04.20.Jb;
05.30.-d; 75.10.Jm

\end{abstract}
\newpage
%%%%%%%%%%%%%%%%%%%%%%%%%%%%%%%%%%%%%%%%%%%%%%%%%%%%%%%%%%%%%%%
%                                                             %
%  1. Introduction                                            %
%                                                             %
%%%%%%%%%%%%%%%%%%%%%%%%%%%%%%%%%%%%%%%%%%%%%%%%%%%%%%%%%%%%%%%
\section{Introduction}
\label{intro} \setcounter{equation}{0}

Gaudin type models constitute a
particularly important class of one-dimensional many-body systems with
long-range interactions. They have found  applications in many
branches of fields ranging from condensed matter physics to high energy
physics. For example, Gaudin models
have been used to establish the integrability of  the reduced
BCS theory  of small
metallic grains \cite{Cam97,Ami01, Del01,Duk04} and
%
%theoretical nuclear
%physics \cite{Dea03,Duk04}, quantum chromo-dynamics (QCD) theory
%\cite{Iac94, Ris00},
the Seiberg-Witten supersymmetric Yang-Mills
theory \cite{Sei94}. They have also provided  a powerful tool for
constructing the solutions to the
Knizhnik-Zamolodchikov equation \cite{Bab93,Has94,Fei94,Hik95,Gou02}
of the Wess-Zumino-Novikov-Witten
conformal
field theory.

%It is well-known that  the periodic Gaudin's magnet Hamiltonians
%(or Gaudin operators) can be constructed via the quasi-classical
%expansion of the transfer matrix ({\it row-to-row transfer
%matrix\/}) of an inhomogeneous spin chain with periodic boundary
%conditions \cite{Hik92,Skl96}. Gaudin models with non-trivial
%boundary conditions can be in principle  treated by means of the
%Sklyanin's boundary inverse scattering method \cite{Skl88}. In
%this case the quasi-classical expansion of the corresponding
%boundary transfer matrix ({\it double-row transfer matrix\/})
%produces {\it generalized\/}

Recently Gaudin models with non-trivial boundaries have attracted
much interest \cite{Hik95, Lor02,Yan041,Yan044,Lim05,Duk01}. So
far, attention has largely been concentrated on Gaudin models with
boundary conditions specified by {\it diagonal\/} K-matrices. In
\cite{Yan041},  the XXZ Gaudin model with boundaries given  by the
{\it non-diagonal\/} K-matrices in \cite{Dev93, Gho94} was
constructed and  solved by the algebraic Bethe ansatz method. In
this paper we generalize the results in \cite{Yan041} and  solve
the $A_{n-1}$ Gaudin magnets with open boundary conditions
corresponding to the {\it non-diagonal\/} K-matrices obtained in
\cite{Yan042}.

This paper is organized as follows. In section 2, we briefly
review the inhomogeneous $A^{(1)}_{n-1}$ trigonometric vertex
model with  integrable boundaries, which also services as
introducing our notation and some basic ingredients . In section
3, we construct the generalized Gaudin operators associated with
non-diagonal K-matrices. The commutativity of these operators
follows from applying the standard procedure
\cite{Hik92,Skl96,Hik95,Lor02}  to the inhomogeneous
$A^{(1)}_{n-1}$ trigonometric vertex  model with off-diagonal
boundaries found in \cite{Yan042}, thus ensuring the integrability
of the Gaudin magnets. In section 4, we diagonalize the Gaudin
operators simultaneously by means of the algebraic Bethe ansatz
method. This constitutes the main new result of this paper. The
diagonalization is achieved by means of the technique of the
``vertex-face" transformation \cite{Yan051}. Section 5 is for
conclusions. In the Apendix, we list the explicit matrix
expressions of the K-matrices corresponding to  the $n=3,4$ cases.

%%%%%%%%%%%%%%%%%%%%%%%%%%%%%%%%%%%%%%%%%%%%%%%%%%%%%%%%%%%%%%%
%                                                             %
%  2. Preliminaries: inhomogeneous $A^{(1)}_{n-1}$            %
%                      open chain                             %
%                                                             %
%                                                             %
%%%%%%%%%%%%%%%%%%%%%%%%%%%%%%%%%%%%%%%%%%%%%%%%%%%%%%%%%%%%%%%

\section{ Preliminaries: inhomogeneous $A^{(1)}_{n-1}$ open chain}
\label{SC} \setcounter{equation}{0} Let us fix a positive integer
$n$ ($n\geq 2$) and a generic complex number $\eta$, and  $R(u)\in
End(\Cb^n\otimes\Cb^n)$ be the  R-matrix of the $A^{(1)}_{n-1}$
trigonometric vertex model given by \cite{Che80,Per81,Baz91} \bea
\hspace{-0.5cm}R(u)=\sum_{\a=1}^{n}R^{\a\a}_{\a\a}(u)E_{\a\a}\otimes
E_{\a\a} +\sum_{\a\ne \b}\lt\{R^{\a\b}_{\a\b}(u)E_{\a\a}\otimes
E_{\b\b}+ R^{\b\a}_{\a\b}(u)E_{\b\a}\otimes
E_{\a\b}\rt\},\label{R-matrix} \eea where $E_{ij}$ is the matrix
with elements $(E_{ij})^l_k=\d_{jk}\d_{il}$. The coefficient
functions are \bea
R^{\a\b}_{\a\b}(u)&=&\lt\{\begin{array}{cc}\frac{\sin(u)}
{\sin(u+\eta)}\,e^{-i\eta},&
\a>\b,\\[6pt]1,&\a=\b,\\[6pt]\frac{\sin(u)}{\sin(u+\eta)}\,e^{i\eta},&
\a<\b,\end{array}\rt.\label{Elements1}\\[6pt]
R^{\b\a}_{\a\b}(u)&=&\lt\{\begin{array}{cc}\frac{\sin(\eta)}{\sin(u+\eta)}\,e^{iu},&
\a>\b,\\[6pt]1,&\a=\b,\\[6pt]\frac{\sin(\eta)}{\sin(u+\eta)}\,e^{-iu},&
\a<\b.\end{array}\rt.\label{Elements2}\eea The R-matrix satisfies
the quantum Yang-Baxter equation (QYBE)
\begin{eqnarray}
 R_{12}(u_1-u_2)R_{13}(u_1-u_3)R_{23}(u_2-u_3)=
 R_{23}(u_2-u_3)R_{13}(u_1-u_3)R_{12}(u_1-u_2), \label{QYB}
\end{eqnarray}
and the properties \cite{Yan042}:\begin{eqnarray}
 &&\hspace{-1.5cm}\mbox{
 Unitarity}:\hspace{42.5mm}R_{12}(u)R_{21}(-u)= {\rm id},\label{Unitarity}\\
 &&\hspace{-1.5cm}\mbox{
 Crossing-unitarity}:\quad
 R^{t_2}_{12}(u)M_2^{-1}R_{21}^{t_2}(-u-n\eta)M_2
 = \frac{\sin(u)\sin(u+n\eta)}{\sin(u+\eta)\sin(u+n\eta-\eta)}\,\mbox{id},
 \label{crosing-unitarity}\\
 &&\hspace{-1.5cm}\mbox{ Quasi-classical
 property}:\hspace{22.5mm}\, R_{12}(u)|_{\eta\rightarrow 0}= {\rm
id}.\label{quasi}
\end{eqnarray}
Here $R_{21}(u)=P_{12}R_{12}(u)P_{12}$ with $P_{12}$ being the
usual permutation operator and $t_i$ denotes the transposition in
the $i$-th space, and $\eta$ is the so-called crossing paramter.
The crossing matrix $M$ is a diagonal $n\times n$ matrix with
elements \bea
M_{\a\b}=M_{\a}\d_{\a\b},~~M_{\a}=e^{-2i\a\eta},~\a=1,\ldots,n.\label{C-Matrix}\eea
Here and below we adopt the standard notation: for any matrix
$A\in {\rm End}(\Cb^n)$, $A_j$ is an embedding operator in the
tensor space $\Cb^n\otimes \Cb^n\otimes\cdots$, which acts as $A$
on the $j$-th space and as an identity on the other factor spaces;
$R_{ij}(u)$ is an embedding operator of R-matrix in the tensor
space, which acts as an identity on the factor spaces except for
the $i$-th and $j$-th ones. The quasi-classical properties
(\ref{quasi}) of the R-matrix enables one to introduce the
corresponding classical r-matrix $r(u)$
\begin{eqnarray}
 R(u)&=&{\rm id} +\eta\,r(u)+O(\eta^2),\qquad\qquad  {\rm
 when}~\eta\longrightarrow 0,\no\\[4pt]
 r(u)&=&\sum_{\a\ne \b}\lt\{r^{\a\b}_{\a\b}(u)E_{\a\a}\otimes
E_{\b\b}+ r^{\b\a}_{\a\b}(u)E_{\b\a}\otimes
E_{\a\b}\rt\}.\label{r-matrix}
\end{eqnarray}
Here the coefficient functions in the expression of $r(u)$ are
given by \bea r^{ij}_{kl}(u)=\lt.\frac{\partial}{\partial
\eta}\lt\{R^{ij}_{kl}(u)\rt\}\rt|_{\eta=0}.\eea

One introduces  the ``row-to-row" monodromy matrix $T(u)$, which
is an $n\times n$ matrix with elements being operators acting  on
$(\Cb^n)^{\otimes N}$  \begin{eqnarray}
T(u)=R_{01}(u+z_1)R_{02}(u+z_2)\cdots
R_{0N}(u+z_N).\label{T-matrix}\end{eqnarray} Here
$\{z_i|i=1,\ldots, N\}$ are arbitrary free complex parameters
which are usually called inhomogeneous parameters. With the help
of the QYBE (\ref{QYB}), one can show that $T(u)$ satisfies the
so-called ``RLL" relation
\begin{eqnarray}
R_{12}(u-v)T_1(u)T_2(v)=T_2(v)T_1(u)R_{12}(u-v).\label{Relation1}\end{eqnarray}

Integrable open chains can be constructed as follows \cite{Skl88}.
Let us introduce a pair of K-matrices $K^-(u)$ and $K^+(u)$. The
former satisfies the reflection equation (RE)
 \begin{eqnarray}
 &&R_{12}(u_1-u_2)K^-_1(u_1)R_{21}(u_1+u_2)K^-_2(u_2)\no\\
  &&~~~~~~=
 K^-_2(u_2)R_{12}(u_1+u_2)K^-_1(u_1)R_{21}(u_1-u_2),\label{RE-V}
\end{eqnarray}
and the latter  satisfies the dual RE \cite{Skl88,Mez91}\bea
&&R_{12}(u_2-u_1)K^+_1(u_1)\,M_1^{-1}\,R_{21}(-u_1-u_2-n\eta)\,M_1\,K^+_2(u_2)\no\\
&&\qquad\quad=
M_1\,K^+_2(u_2)R_{12}(-u_1-u_2-n\eta)\,M_1^{-1}\,K^{+}_1(u_1)R_{21}(u_2-u_1).
\label{DRE-V1}\eea

For models with open boundaries, instead of the standard
``row-to-row" monodromy matrix $T(u)$ (\ref{T-matrix}), one needs
the  ``double-row" monodromy matrix $\mathbb{T}(u)$
\begin{eqnarray}
 \mathbb{T}(u)=T(u)K^-(u)T^{-1}(-u).\label{Mon-V-1}
\end{eqnarray}
Using (\ref{Relation1}) and (\ref{RE-V}), one can prove that
$\mathbb{T}(u)$ satisfies
\begin{eqnarray}
 R_{12}(u_1-u_2)\mathbb{T}_1(u_1)R_{21}(u_1+u_2)
  \mathbb{T}_2(u_2)=
 \mathbb{T}_2(u_2)R_{12}(u_1+u_2)\mathbb{T}_1(u_1)R_{21}(u_1-u_2).
 \label{Relation-Re}
\end{eqnarray}
Then the {\it double-row transfer matrix\/} of  the inhomogeneous
$A^{(1)}_{n-1}$ trigonometric vertex model  with open boundary is
given by
\begin{eqnarray}
\t(u)=tr(K^+(u)\mathbb{T}(u)).\label{trans}
\end{eqnarray}
The commutativity of the transfer matrices
\begin{eqnarray}
 [\t(u),\t(v)]=0,\label{Com-2}
\end{eqnarray}
follows as a consequence of (\ref{QYB})-(\ref{crosing-unitarity})
and (\ref{RE-V})-(\ref{DRE-V1}). This ensures the integrability of
the inhomogeneous $A^{(1)}_{n-1}$ trigonometric vertex models with
open boundaries specified by the K-matrices.

%%%%%%%%%%%%%%%%%%%%%%%%%%%%%%%%%%%%%%%%%%%%%%%%%%%%%%%%%%%%%%%
%                                                             %
%  3. $A_{n-1}$ Gaudin models with  boundaries                %
%                                                             %
%                                                             %
%%%%%%%%%%%%%%%%%%%%%%%%%%%%%%%%%%%%%%%%%%%%%%%%%%%%%%%%%%%%%%%
\section{$A_{n-1}$ Gaudin model with  boundaries}
 \label{Haml} \setcounter{equation}{0}
In this paper, we will consider  the non-diagonal solutions
$K^{\pm}(u)$ obtained in  \cite{Yan042}, which are respectively
given by
 \bea K^-(u)^s_t&=&\sum_{j=1}^n
k_j(u)\phi^{(s)}_j(u;\l)
\bar{\phi}^{(t)}_j(-u;\l), \label{K-matrix}\\
K^+(u)^s_t&=&\sum_{j=1}^n
\tilde{k}_j(u)\phi^{(s)}_j(-u;\l')
\tilde{\phi}^{(t)}_j(u;\l'). \label{DK-matrix}\eea
There are $n$ different solutions, each parametrized by an integer $l$
($1\leq l\leq n$) and
%The functions  $\{k_j(u)|j=1,\ldots,n\}$ and
%$\{\tilde{k}_j(u)|j=1,\ldots,n\}$ are
given by
\bea k_j(u)&=&\lt\{\begin{array}{ll}1,&\qquad\quad\quad 1\leq j\leq
l,\\[4pt]
\frac{\sin(\xi-u)}{\sin(\xi+u)}e^{-2iu},&\qquad\quad\quad l+1\leq j\leq
n,\end{array}\rt.\label{K-matrix1}\\
\tilde{k}_j(u)&=&\lt\{\begin{array}{ll}e^{-2i(j\eta)},&1\leq j\leq l,\\[4pt]
\frac{\sin(\bar{\xi}+u+\frac{n}{2}\eta)}{\sin(\bar{\xi}-u-\frac{n}{2}\eta)}
e^{2i(u+\frac{n-2j}{2}\eta)},&l+1\leq j\leq
n.\end{array}\rt.\label{DK-matrix1}\eea  Hereafter we choose some
particular $l$ without losing generality.
In (\ref{K-matrix}) and
(\ref{DK-matrix}), $\phi$, $\bar{\phi}$ and $\tilde{\phi}$ are
intertwiners which are specified in section 4. Besides a {\it
discrete\/} parameter $l$, the K-matrix $K^-(u)$ (resp. $K^+(u)$)
depends on  {\it continuous\/} parameters $\xi$, $\{\l_j\}$
(resp. $\bar{\xi}$, $\{\l'_j\}$) and $\rho$ (whose dependence is
through the intertwiner-matrix (\ref{In-matrix})
below). Here and throughout, associated with
the boundary parameters $\{\l_j\}$ (resp. $\{\l'_j\}$) we
introduce a vector $\l=\sum_{j=1}^n\l_j\e_j $ (resp.
$\l'=\sum_{j=1}^n\l'_j\e_j$), where $\{\e_j|\,j=1,\ldots,n\}$ is
the orthonormal basis of the vector space $\Cb^n$ such that
$\langle \e_j,\e_k\rangle=\d_{jk}$.

As can be seen below (e.g.  (\ref{In-matrix}), (\ref{Int1}) and
(\ref{Int2})), when $m$ is specialized to $\l$, $K^-(u)$ does not depend
on
$\eta$. Moreover,  the dependence on $\l_n$  {\it
disappears\/}  in the final expression of
$K^-(u)$ although it appears in the expression of
$\phi_{i}(u;\l)$. Without loss of generality in this paper
we will assume $\l_n=0$.  Thus the K-matrix
$K^-(u)$  depends on $n+1$ continuous free parameters
 $\{\l_j|j=1,\ldots,n-1;\,\rho,\,\xi\}$ for $1\leq l\leq n-1$,
 and $n$ parameters $\{\l_j|j=1,\ldots,n-1;\,\rho\}$ for $l=n$.
Some explicit matrix forms of the K-matrices $K^-(u)$ for the
$n=3,4$ cases are given as (\ref{K1-1})-(\ref{K2-2}) in the
Appendix A.

Let us emphasize  that a further restriction \bea
\l'+\eta\sum_{k=1}^N\e_{j_k}=\l,\label{Restriction}\eea where
$\{j_k|k=1,\ldots,N\}$ are positive integers such that $ 2\leq
j_k\leq n$, is necessary for  the application of the algebraic
Bethe ansatz method in section 4. Hereafter, we shall impose this
constraint. We further restrict the complex parameters $\xi$ and
$\bar{\xi}$ to be the same, i.e.,\begin{eqnarray}
\bar{\xi}=\xi,\label{Restriction-1}\end{eqnarray} so that
(\ref{ID-1}) below is satisfied. Under these two constraits,
the K-matrix $K^{+}(u)$
depends on the same set of free parameters as the K-matrix $K^-(u)$ (see
above). Moreover, the
K-matrices satisfy the following relation thanks to the
restrictions (\ref{Restriction}) and (\ref{Restriction-1})
\begin{eqnarray}
\lim_{\eta\rightarrow 0}\{K^+(u)\,K^-(u)\}=\lim_{\eta\rightarrow
0}\{K^+(u)\}K^-(u)={\rm id}.\label{ID-1}
\end{eqnarray}

Let us now introduce the  Gaudin operators $\{H_j
|j=1,2,\cdots,N\}$ associated with the inhomogeneous
$A^{(1)}_{n-1}$ trigonometric vertex model with  boundaries
specified by the K-matrices (\ref{K-matrix}) and
(\ref{DK-matrix}):
\begin{eqnarray}
H_j=\G_j(z_j)+\sum_{k\neq
 j}^{N}r_{kj}(z_j-z_k)+\lt(K^{-}_j(z_j)\rt)^{-1}\lt\{\sum_{k\neq
 j}^{N}r_{jk}(z_j+z_k)\rt\}K^-_j(z_j),\label{Ham}
\end{eqnarray}
where $\G_j(u)=\frac{\partial}{\partial
\eta}\{\bar{K}_j(u)\}|_{\eta=0}K^-_j(u)$, $j=1,\cdots,N,$ with
$\bar{K}_j(u)=tr_0\lt\{K^+_0(u)R_{0j}(2u)P_{0j}\rt\}$. Here
$\{z_j\}$ are the inhomogeneous parameters of the inhomogeneous
$A^{(1)}_{n-1}$ trigonometric vertex  model and $r(u)$ is the
classical r-matrix given by (\ref{r-matrix}).

The generalized Gaudin operators (\ref{Ham}) are obtained by
expanding the double-row transfer matrix (\ref{trans}) at the
point $u=z_j$ around $\eta=0$:
\begin{eqnarray}
 \t(z_j)&=&\t(z_j)|_{\eta=0}+\eta
 H_j+O(\eta^2),\quad j=1,\cdots,N, \label{trans-2}\\
 H_j&=&\frac{\partial}{\partial \eta}\t(z_j)|_{\eta=0}\,  .\label{Eq-1}
\end{eqnarray}
The relations (\ref{quasi}) and (\ref{ID-1}) imply that the first
term in the expansion (\ref{trans-2}) is equal to the identity,
\begin{eqnarray}
 \t(z_j)|_{\eta=0}={\rm id}.\label{First}
\end{eqnarray}
Then the commutativity of the transfer matrices $\{\t(z_j)\}$
(\ref{Com-2})  implies
\begin{eqnarray}
[H_j,H_k]=0,\quad j,k=1,\cdots,N.\label{Com-1}
\end{eqnarray} Thus the
Gaudin system defined by (\ref{Ham}) is integrable. Moreover, the
fact that the Gaudin operators $\{H_j\}$ (\ref{Ham}) can be
expressed in terms of the transfer matrix of the inhomogeneous
$A^{(1)}_{n-1}$ trigonometric vertex model with open boundary
enables us to exactly diagonalize the operators by the algebraic
Bethe ansatz method with the help of the ``vertex-face"
correspondence technique, as will be shown in the next section.
The aim of this paper is to diagonalize  the generalized Gaudin
operators $H_j$, $j=1,\cdots,N$, (\ref{Ham}), simultaneously.

%%%%%%%%%%%%%%%%%%%%%%%%%%%%%%%%%%%%%%%%%%%%%%%%%%%%%%%%%%%%%%%
%                                                             %
%  4. Eigenvalues and Bethe ansatz equations                  %
%                                                             %
%                                                             %
%%%%%%%%%%%%%%%%%%%%%%%%%%%%%%%%%%%%%%%%%%%%%%%%%%%%%%%%%%%%%%%

\section{Eigenvalues and Bethe ansatz equations}
\label{BAE} \setcounter{equation}{0}
\subsection{Intertwining vectors and face-vertex
correspondence}

For a vector $m\in \Cb^n$, set \bea m_j=\langle m,\e_j\rangle,
~~|m|=\sum_{k=1}^nm_k,\quad j=1,\ldots,n. \label{Def1}\eea

We introduce \cite{Yan042} $n$ intertwining vectors
(intertwiners) $\{\phi_j(u;m)|\, j=1,\ldots,n\}$. Each
$\phi_j(u;m)$ is an $n$-component column vector whose
$\a$-th component is $\{\phi^{(\a)}_j(u;m)\}$. The $n$
intertwiners form an $n\times n$ the intertwiner-matrix (in which
$j$ and $\a$ stand for the column and the row indices
respectively),  with the non-vanishing matrix elements being \bea
\lt(\begin{array}{ccccccc}e^{i f_1(m)} &&&&&&e^{i
F_n(m)+\rho}e^{2iu}\\e^{i F_1(m)}&e^{i f_2(m)}&&&&&\\
&e^{i F_2(m)}&\ddots&&&&\\&&\ddots&e^{i f_j(m)}&&&\\&&&e^{i
F_j(m)}&\ddots&&\\&&&&\ddots&e^{i f_{n-1}(m)}&\\&&&&&e^{i
F_{n-1}(m)}&e^{i f_n(m)}
\end{array}\rt).\label{In-matrix}\eea Here $\rho$  is a complex constant,
and $\{f_j(m)|j=1,\ldots,n\}$ and
$\{F_{j}(m)|j=1,\ldots,n\}$ are linear functions of $m$:\bea
f_j(m)&=&\sum_{k=1}^{j-1}m_k-m_j-\frac{1}{2}|m|,~~j=1,\ldots,n,\label{function1}\\
F_{j}(m)&=&\sum_{k=1}^{j}m_k-\frac{1}{2}|m|,~~j=1,\ldots,n-1,\label{function2}\\
F_n(m)&=&-\frac{3}{2}|m|.\label{function3}\eea We remark that as is clear
from (\ref{In-matrix}), while $\phi_n(u;m)$ is a function of
$u$, $\phi_j(u;m)$ does not depend on $u$ for $1\leq
j\leq n-1$.

{}From the above
intertwiner-matrix, one may derive  the following {\it face-vertex
correspondence relation\/} \cite{Yan042}:\bea &&R_{12}(u_1-u_2)
\phi_i(u_1;m)\otimes
\phi_j(u_2;m-\eta\e_i) \no\\
&&~~~~=\sum_{k,l}W^{kl}_{ij}(u_1-u_2)
\phi_k(u_1;m-\eta\e_l)\otimes
\phi_l(u_2;m).\label{Face-vertex}\eea Here the
non-vanishing elements of $\{W^{kl}_{ij}(u)\}$ are \bea
W^{jj}_{jj}(u)&=&1,~~W^{jk}_{jk}(u)=\frac{\sin(u)}{\sin(u+\eta)},
~~{\rm for}\,\,j\neq k,\label{W-elements-1}\\[6pt]
W^{kj}_{jk}(u)&=&\lt\{\begin{array}{cc}\frac{\sin(\eta)}
{\sin(u+\eta)}\,e^{iu},&
j>k,\\[6pt]\frac{\sin(\eta)}{\sin(u+\eta)}\,e^{-iu},&
j<k,\end{array}\rt.~\quad {\rm for}~j\neq k.\label{W-elements-2}\eea
Associated with $\{W^{kl}_{ij}(u)\}$, one may introduce ``{\it
face"\/} type R-matrix $W(u)$ \bea W(u)=\sum_{i,j,k,l}
W^{kl}_{ij}(u)E_{ki}\otimes E_{lj}.\label{W-matrix}\eea Note that
the ``face" type R-matrix $W(u)$ does not depend on the face type
parameter $m$, in contrast to
 the $\Zb_n$ elliptic case \cite{Jim87}, and thus
 $W(u)$ and $R(u)$
 satisfy the same QYBE, i.e. $W(u)$ obeys the usual (vertex type) QYBE rather than
 the {\it dynamical\/} one. Moreover the R-matrix $W(u)$
 enjoys  the quasi-classical property, \bea W(u)|_{\eta\rightarrow 0}={\rm
 id}.\label{W-1}\eea
For a generic $\rho\in\Cb$, the determinant of the intertwiner matrix
(\ref{In-matrix}) is non-vanishing  and thus the inverse of
(\ref{In-matrix}) exists \cite{Yan042}. This fact allows the introducation
of other
types of intertwiners $\bar{\phi}$ and $\tilde{\phi}$ satisfying
the following orthogonality conditions: \bea
&&\sum_{\a}\bar{\phi}^{(\a)}_{i}(u;m)
~\phi^{(\a)}_{j}(u;m)=\d_{ij},\label{Int1}\\[6pt]
&&\sum_{\a}\tilde{\phi}^{(\a)}_{i}(u;m+\eta\e_i)
~\phi^{(\a)}_{j}(u;m+\eta\e_j)=\d_{ij}.\label{Int2}\eea From these
conditions we  derive the ``completeness" relations:\bea
&&\sum_{k}\bar{\phi}^{(\a)}_{k}(u;m)
~\phi^{(\b)}_{k}(u;m)=\d_{\a\b},\label{Int3}\\[6pt]
&&\sum_{k}\tilde{\phi}^{(\a)}_{k}(u;m+\eta\e_k)
~\phi^{(\b)}_{k}(u;m+\eta\e_k)=\d_{\a\b}.\label{Int4}\eea

%With the help of (\ref{Int1})-(\ref{Int4}), we  obtain, from the
%face-vertex correspondence relation (\ref{Face-vertex}),
%\begin{eqnarray}
% &&\left(\tilde{\phi}_{m+\eta\e_{k},m}(u_1)\otimes
% {\rm id}\right)R_{12}(u_1-u_2) \left({\rm
%id}\otimes\phi_{m+\eta\e_{j},m}(u_2)\right)\no\\
% &&\qquad\quad= \sum_{i,l}W^{kl}_{ij}(u_1-u_2)\,
% \tilde{\phi}_{m+\eta(\e_{i}+\e_{j}),m+\eta\e_{j}}(u_1)\otimes
% \phi_{m+\eta(\e_{k}+\e_{l}),m+\eta\e_{k}}(u_2),\label{Face-vertex1}\\
% &&\left(\tilde{\phi}_{m+\eta\e_{k},m}(u_1)\otimes
%
%\tilde{\phi}_{m+\eta(\e_{k}+\e_{l}),m+\eta\e_{k}}(u_2)\right)R_{12}(u_1-u_2)\no\\
% &&\qquad\quad= \sum_{i,j}W^{kl}_{ij}(u_1-u_2)\,
% \tilde{\phi}_{m+\eta(\e_{i}+\e_{j}),m+\eta\e_{j}}(u_1)\otimes
% \tilde{\phi}_{m+\eta\e_{j},m}(u_2),\label{Face-vertex2}\\
% &&\left({\rm id}\otimes
% \bar{\phi}_{m,m-\eta\e_{l}}(u_2)\right)R_{12}(u_1-u_2)
% \left(\phi_{m,m-\eta\e_{i}}(u_1)\otimes {\rm id}\right)\no\\
% &&\qquad\quad= \sum_{k,j}W^{kl}_{ij}(u_1-u_2)\,
% \phi_{m-\eta\e_{l},m-\eta(\e_{k}+\e_{l})}(u_1)\otimes
%
%\bar{\phi}_{m-\eta\e_{i},m-\eta(\e_{i}+\e_{j})}(u_2),\label{Face-vertex3}\\
% &&\left(\bar{\phi}_{m-\eta\e_{l},m-\eta(\e_{k}+\e_{l})}(u_1)\otimes
% \bar{\phi}_{m,m-\eta\e_{l}}(u_2)\right)R_{12}(u_1-u_2)\no\\
% &&\qquad\quad= \sum_{i,j}W^{kl}_{ij}(u_1-u_2)\,
% \bar{\phi}_{m,m-\eta\e_{i}}(u_1)\otimes
% \bar{\phi}_{m-\eta\e_{i},m-\eta(\e_{i}
% +\e_{j})}(u_2).\label{Face-vertex4}
%\end{eqnarray}
Corresponding to the vertex type K-matrices (\ref{K-matrix}) and
(\ref{DK-matrix}), one has the following face type
K-matrices $\K$ and $\tilde{\K}$ \cite{Yan03}
\begin{eqnarray}
 &&\K(\l|u)^j_i=\sum_{\a,\b}\tilde{\phi}^{(\a)}
 _{j}(u;\l-\eta(\e_i-\e_j))\,K^-(u)^{\a}_{\b}\,\phi^{(\b)}
 _{i}(-u;\l),\label{K-F-1}\\
 &&\tilde{\K}(\l'|u)^j_i=\sum_{\a,\b}\bar{\phi}^{(\a)}
 _{j}(-u;\l')\,K^+(u)^{\a}_{\b}\,\phi^{(\b)}
 _{i}(u;\l'-\eta(\e_j-\e_i)).\label{K-F-2}
\end{eqnarray}
Straightforward calculations show that the face type
K-matrices  are  {\it diagonal\/} \footnote{The spectral parameter
$u$ and the boundary parameter $\xi$ of the reduced double-row
monodromy matrices constructed from $\K(\l|u)$ will be shifted in
each step of the nested Bethe ansatz procedure \cite{Yan04}.
Therefore, it is convenient to specify the dependence of
$\K(\l|u)$ on the boundary parameter $\xi$ through $k_j(u;\xi)$.}
\begin{eqnarray}
\K(\l|u)^j_i=\d_i^j\,k_j(u;\xi),\quad
\tilde{\K}(\l'|u)^j_i=\d_i^j\,\tilde{k}_j(u),~~\qquad i,j=1,\ldots,n,
\label{Diag-F}
\end{eqnarray}
where functions $\{k_i(u;\xi)=k_i(u)\}$ and $\{\tilde{k}_i(u) \}$
are respectively given by (\ref{K-matrix1}) and
(\ref{DK-matrix1}).

A remark is in order. Although the K-matrices $K^{\pm}(u)$ given
by (\ref{K-matrix}) and (\ref{DK-matrix}) are generally
non-diagonal (in the vertex picture), after the face-vertex
transformations (\ref{K-F-1}) and (\ref{K-F-2}), the face type
counterparts $\K(\l|u)$ and $\tilde{\K}(\l'|u)$ become diagonal
{\it simultaneously\/}. This fact enables ones  to apply the
generalized algebraic Bethe ansatz method to diagonalize the
transfer matrix $\tau(u)$ \cite{Cao03,Yan041,Yan051} (for $n=2$
case, the corresponding transfer matrix was  diagonalized
alternatively by the fusion hierarchy of the transfer matrices
with the anisotropy value being the roots of unity \cite{Nep03,Nep04}).

\subsection{Algebraic Bethe ansatz}
By means of (\ref{Int3}), (\ref{Int4}), (\ref{K-F-2}) and
(\ref{Diag-F}), the transfer matrix $\t(u)$ (\ref{trans}) can be
recasted  into the following face type form:
\begin{eqnarray}
 \t(u)
%&=&tr(K^+(u)\mathbb{T}(u))\no\\
% &=&\sum_{\mu,\nu}tr\lt(\!K^+(u)\, \phi_{\l'-\eta(\e_{\mu}-\e_{\nu}),
% \l'-\eta\e_{\mu}}(u)\,\tilde{\phi}_{\l'-\eta(\e_{\mu}-\e_{\nu}),
% \l'-\eta\e_{\mu}}(u)~\rt.\no\\
% &&\qquad\times\lt.\mathbb{T}(u) \phi_{\l', \l'-\eta\e_{\mu}}(-u)
% \bar{\phi}_{\l', \l'-\eta\e_{\mu}}(-u)\rt)\no\\
% &=&\sum_{\mu,\nu}\bar{\phi}_{\l', \l'-\eta\e_{\mu}}(-u)K^+(u)
% \,\phi_{\l'-\eta(\e_{\mu}-\e_{\nu}),\l'-\eta\e_{\mu}}(u)\no\\
%&&\qquad\times\tilde{\phi}_{\l'-\eta(\e_{\mu}-\e_{\nu}),
% \l'-\eta\e_{\mu}}(u)~\mathbb{T}(u)\phi_{\l', \l'-\eta\e_{\mu}}(-u)\no\\
=\sum_{\mu,\nu}\tilde{\K}(\l'|u)_{\nu}^{\mu}\,\T(\l'|u)^{\nu}_{\mu}=
 \sum_{\mu}\tilde{k}_{\mu}(u)\,\T(\l'|u)^{\mu}_{\mu}.
 \label{De1}
\end{eqnarray}
Here we have introduced the face type double-row monodromy matrix
$\T(m|u)$,
\begin{eqnarray}
 \T(\l'|u)^{\nu}_{\mu}&=&\lt.\lt.\T(m|u)^{\nu}_{\mu}\rt|_{m=\l'}=
 \tilde{\phi}_{\nu}
(u;m-\eta(\e_{\mu}-\e_{\nu}))~\mathbb{T}(u)\,\phi_{\mu}(-u;m)
\rt|_{m=\l'}\no\\
 &\equiv&
 \sum_{\a,\b}\tilde{\phi}^{(\b)}
_{\nu}(u;\l'-\eta(\e_{\mu}-\e_{\nu}))
~\mathbb{T}(u)^{\b}_{\a}\,\phi^{(\a)}_{\mu}(-u;\l').\label{Mon-F}
\end{eqnarray}
Moreover, (\ref{Relation-Re}), (\ref{Face-vertex}) and
(\ref{Int4}) imply the following exchange relations among
$\T(m|u)^{\nu}_{\mu}$:
\begin{eqnarray}
 &&\sum_{i_1,i_2}\sum_{j_1,j_2}~
 W^{i_0\,j_0}_{i_1\,j_1}(u_1-u_2)\,\T(m+\eta(\e_{j_1}+\e_{i_2})|u_1)
 ^{i_1}_{i_2}\no\\
 &&~~~~~~~~\times W^{j_1\,i_2}_{j_2\,i_3}(u_1+u_2)\,
 \T(m+\eta(\e_{j_3}+\e_{i_3})|u_2)^{j_2}_{j_3}\no\\[2pt]
 &&~~=\sum_{i_1,i_2}\sum_{j_1,j_2}~
 \T(m+\eta(\e_{j_1}+\e_{i_0})|u_2)
 ^{j_0}_{j_1}\,W^{i_0\,j_1}_{i_1\,j_2}(u_1+u_2)\no\\
 &&~~~~~~~~\times\T(m+\eta(\e_{j_2}+\e_{i_2})|u_1)^{i_1}_{i_2}\,
 W^{j_2\,i_2}_{j_3\,i_3}(u_1-u_2).\label{RE-F}
\end{eqnarray}

In the following we will use the standard
notation,
\begin{eqnarray}
\A(m|u)&=&\hspace{-0.3cm}\T(m|u)^1_1,~\B_j(m|u)=\hspace{-0.1cm}
\T(m|u)^1_j,~\C_j(m|u)=\hspace{-0.1cm}
\T(m|u)^j_1,~j=2,\ldots,n,\label{Def-AB}\\
\D^j_i(m|u)&=&\hspace{-0.3cm}\T(m|u)^j_i-\d^j_iW^{j\,1}_{1\,j}(2u)\,\A(m|u),
\qquad\qquad\qquad\qquad\quad
 i,j=2,\ldots,n. \label{Def-D}
\end{eqnarray}
In order to apply the algebraic Bethe ansatz method,
one needs to construct a pseudo-vacuum state (also called a
reference state) which is the common eigenstate of the operators
$\A$, $\D^j_j$ and is annihilated by the operators $\C_j$. In
contrast to the models with {\it diagonal\/} $K^{\pm}(u)$,
the usual highest-weight state
\begin{eqnarray}
 \lt(\begin{array}{l}1\\0\\\vdots\end{array}\rt)\otimes\cdots\otimes
 \lt(\begin{array}{l}1\\0\\\vdots\end{array}\rt),\no
\end{eqnarray}
is no longer the pseudo-vacuum state. However, after the
face-vertex transformations (\ref{K-F-1}) and (\ref{K-F-2}), the
face type K-matrices $\K(\l|u)$ and $\tilde{\K}(\l|u)$ {\it
simultaneously\/} become diagonal. This suggests that one can
translate the $A^{(1)}_{n-1}$ trigonometric  model  with
non-diagonal K-matrices into the corresponding SOS model with {\it
diagonal} K-matrices $\K(\l|u)$ and $\tilde{\K}(\l|u)$ given by
(\ref{K-F-1})-(\ref{K-F-2}).

Consider the  state \cite{Yan051},
\begin{eqnarray}
 |\O\rangle=\phi_1(-z_1;\l-(N-1)\eta\e_{1})\otimes
 \phi_1(-z_{2};\l-(N-2)\eta\e_{1})\cdots\otimes
 \phi_1(-z_N;\l),\label{Vac}
\end{eqnarray}
which  depends  on the boundary parameters $\{\l_j\}$, but not on
the boundary parameter $\xi$. It is also independent of $\{z_i\}$ because
as mentioned before the vector $\phi_1(u;m)$ does not depend on $u$
regardless of  $m$. The state (\ref{Vac}) is  the
pseudo-vacuum state
in the ``face picture" since it can be shown to satisfy the following
equations: \bea \A(\l-N\eta\e_{1}|u)\,|\O\rangle&=&k_1(u;\xi)
|\O\rangle,\label{A}\\
\D^a_j(\l-N\eta\e_{1}|u)\,|\O\rangle&=&\d^a_j
\frac{\sin(2u)e^{i\eta}}{\sin(2u+\eta)}
k_j(u+\frac{\eta}{2};\xi-\frac{\eta}{2})\no\\
&&\qquad\times\lt\{\prod_{k=1}^N
\frac{\sin(u+z_k)\sin(u-z_k)}{\sin(u+z_k+\eta)\sin(u-z_k+\eta)}\rt\}
\,|\O\rangle,\no\\
&&\qquad\qquad\qquad a,j=2,\ldots,n,\label{D}\\
\C_j(\l-N\eta\e_{1}|u)\,|\O\rangle&=&0,~~~\qquad\qquad
j=2,\ldots,n,\\
\B_j(\l-N\eta\e_{1}|u)\,|\O\rangle&\neq& 0,~~~\qquad\qquad
j=2,\ldots,n, \eea
as required.

For later convenience, we introduce a set of non-negative  integers
$\{N_j|j=1,\ldots,n-1\}$ with  $N_1=N$ and complex parameters
$\{v^{(j)}_k|~k=1,2,\ldots,N_{j+1},~j=0,1,\ldots,n-2\}$. As in
\cite{Bab82,Sch83,Dev94,Yan051},
the parameters $\{v^{(j)}_k\}$ will be used to specify the
eigenvectors of the corresponding reduced transfer matrices (see below).
We will also adopt the  convention,
\bea  v_k=v^{(0)}_k,~\qquad\qquad k=1,2,\ldots, N.\eea

We now apply the generalized algebraic Bethe
ansatz method  developed in \cite{Yan04} to diagonalize
the transfer matrix (\ref{trans}) with
the K-matrices $K^{\pm}(u)$ given by
(\ref{K-matrix})-(\ref{Restriction}).
We seek the common eigenvectors of the transfer matrix of the form
\begin{eqnarray}
&&|v_1,\ldots,v_{N}\rangle =\sum_{i_1,\ldots,i_{N}=2}^n
F^{i_1,i_2,\ldots,i_{N}}\,
\B_{i_1}(\l'+\eta\e_{i_1}-\eta\e_{1}|v_{1})\,\B_{i_2}
(\l'+\eta(\e_{i_1}+\e_{i_2})-2\eta\e_{1}|v_2)\cdots\no\\
&&\qquad\qquad\qquad\qquad\qquad\times \B_{i_{N-1}}
(\l'+\eta\sum_{k=1}^{N-1}\e_{i_k}-(N-1)\eta\e_{1}|v_{N-1})\no\\
&&\qquad\qquad\qquad\qquad\qquad\times\B_{i_{N}}
(\l'+\eta\sum_{k=1}^{N}\e_{i_k}-N\eta\e_{1}|v_{N})\,
|\O\rangle.\label{Eigenstate1}
\end{eqnarray}
Here the summing indices should obey the constraint, $
\l'+\eta\sum_{k=1}^N\e_{i_k}=\l$, where
$\l'$ and $\l$ are the boundary parameters which satisfy the
restriction (\ref{Restriction}). Taking this constraint into account,
(\ref{Eigenstate1}) may be written as
\begin{eqnarray}
&&|v_1,\ldots,v_{N}\rangle =\sum_{i_1,\ldots,i_{N}=2}^n
F^{i_1,i_2,\ldots,i_{N}}\,
\B_{i_1}(\l'+\eta\e_{i_1}-\eta\e_{1}|v_{1})\,\B_{i_2}
(\l'+\eta(\e_{i_1}+\e_{i_2})-2\eta\e_{1}|v_2)\cdots\no\\
&&\qquad\qquad\qquad\qquad\qquad\times \B_{i_{N-1}}
(\l'+\eta\sum_{k=1}^{N-1}\e_{i_k}-(N-1)\eta\e_{1}|v_{N-1})\no\\
&& \qquad\qquad\qquad\qquad\qquad\times \B_{i_{N}}
(\l-N\eta\e_{1}|v_{N})\,|\O\rangle.\label{Eigenstate}
\end{eqnarray}

With the help of (\ref{De1}), (\ref{Def-AB}) and (\ref{Def-D}) one may
rewrite the transfer matrix (\ref{trans}) in terms of the
operators $\A$ and $\D^j_j$
\begin{eqnarray}
 \t(u)=\a^{(1)}(u)\,\A(\l'|u)+\sum_{j=2}^n
 \tilde{k}_j^{(1)}(u+\frac{\eta}{2})\,\D(\l'|u)^j_j.
 \label{trans1}
\end{eqnarray}
Here we have introduced the function $\a^{(1)}(u)$,
\begin{eqnarray}
 \a^{(1)}(u)=\sum_{j=1}^n\tilde{k}_j(u)\,
 W^{j\,1}_{\,1j}(2u),\label{function-a}
\end{eqnarray}
and the reduced
 K-matrix $\tilde{\K}^{(1)}(\l'|u)$ with the elements given by
\begin{eqnarray}
\tilde{\K}^{(1)}(\l'|u)^j_i&=&
 \d^j_i\,\tilde{k}_j^{(1)}(u),\qquad\qquad\qquad
 i,j=2,\ldots,n,\label{Reduced-K1}\\
\tilde{k}_j^{(1)}(u)&=& \tilde{k}_j(u-\frac{\eta}{2}),
\qquad\qquad\qquad
 j=2,\ldots,n.
 \label{Reduced-K2}
\end{eqnarray}
Following \cite{Yan04,Yan044,Yan051},
we introduce a set of
reduced K-matrices $\{\tilde{\K}^{(b)}(\l'|u)|b=0,\ldots,n-1\}$  which
include the original one
$\tilde{\K}(\l'|u)=\tilde{\K}^{(0)}(\l'|u)$ and the ones in
(\ref{Reduced-K1}) and (\ref{Reduced-K2}):
\begin{eqnarray}
 \tilde{\K}^{(b)}(\l'|u)^j_i&=&
 \d^j_i\,\tilde{k}_j^{(b)}(u),\quad i,j=b+1,\ldots,n,
 \quad b=0,\ldots,n-1,
 \label{Reduced-K3}\\
 \tilde{k}_j^{(b)}(u)&=&
\tilde{k}_j(u-b\frac{\eta}{2}),\quad j=b+1,\ldots,n,
 \quad b=0,\ldots,n-1.
 \label{Reduced-K4}
\end{eqnarray}
Moreover we introduce a set of functions
$\{\a^{(b)}(u)|b=1,\ldots,n-1\}$ (including the one in
(\ref{function-a})) related to the reduced K-matrices
$\tilde{\K}^{(b)}(\l'|u)$
\begin{eqnarray}
 \a^{(b)}(u)=\sum_{j=b}^{n}W^{j\,b}_{\,bj}(2u)\,
 \tilde{k}_j^{(b-1)}(u),\quad\qquad\qquad\qquad
b=1,\ldots,n.\label{function-a-1}
\end{eqnarray}

Carrying out the nested Bethe ansatz \cite{Yan051}, one
finds that, with the coefficients $F^{i_1,i_2,\cdots,i_{N}}$ in
(\ref{Eigenstate}) properly chosen, the Bethe state
$|v_1,\ldots,v_{N}\rangle $ is the eigenstate of the transfer
matrix (\ref{trans}),
\begin{eqnarray}
 \t(u)\,|v_1,\ldots,v_{N}\rangle=\L(u;\xi,\{v_k\})\,
 |v_1,\ldots,v_{N}\rangle,\end{eqnarray} with the eigenvalue given by
 \begin{eqnarray}
 &&\L(u;\xi,\{v_k\})\no\\[4pt]
 &&\qquad=\a^{(1)}(u)k_1(u;\xi)\,
 \prod_{k=1}^{N_1}\frac{\sin(u+v_k)\sin(u-v_k-\eta)}
 {\sin(u+v_k+\eta)\sin(u-v_k)}\no\\[4pt]
 &&\qquad\quad+\frac{\sin(2u)e^{i\eta}} {\sin(2u+\eta)}\,
\lt\{ \prod_{k=1}^{N_1}\frac{\sin(u-v_k+\eta)\sin(u+v_k+2\eta)}
 {\sin(u-v_k)\sin(u+v_k+\eta)}\rt.\no\\[4pt]
 &&\qquad\qquad\qquad
 \times \prod_{k=1}^{N}\frac{\sin(u+z_k)\sin(u-z_k)}
 {\sin(u+z_k+\eta)\sin(u-z_k+\eta)}\no\\[4pt]
&&\qquad\qquad\qquad\times
\lt.\L^{(1)}(u+\frac{\eta}{2};\xi-\!\frac{\eta}{2},\{v^{(1)}_k\})\rt\}.
\label{Eigenvalue1}
\end{eqnarray}
The eigenvalues
$\{\L^{(j)}(u;\xi,\{v^{(j)}_{k}\})|j=0,\ldots,n-1\}$ (with
$\L(u;\xi,\{v_{k}\})=\L^{(0)}(u;\xi,\{v^{(0)}_{k}\})$) of the
reduced transfer matrices are given by the following recurrence
relations:
\begin{eqnarray}
&& \L^{(j)}(u;\xi^{(j)},\{v^{(j)}_k\})\no\\[4pt]
&&\qquad=\a^{(j+1)}(u)
 k_{j+1}(u;\xi^{(j)})\,
 \prod_{k=1}^{N_{j+1}}\frac{\sin(u+v^{(j)}_k)\sin(u-v^{(j)}_k-\eta)}
 {\sin(u+v^{(j)}_k+\eta)\sin(u-v^{(j)}_k)}\no\\[4pt]
 &&\qquad\quad+\frac{\sin(2u)e^{i\eta}}
 {\sin(2u+\eta)}\,
 \lt\{\prod_{k=1}^{N_{j+1}}\frac{\sin(u-v^{(j)}_k+\eta)\sin(u+v^{(j)}_k+2\eta)}
 {\sin(u-v^{(j)}_k)\sin(u+v^{(j)}_k+\eta)}\rt.\no\\[4pt]
 &&\qquad\qquad\qquad\times
 \prod_{k=1}^{N_j}\frac{\sin(u+z^{(j)}_k)\sin(u-z^{(j)}_k)}
 {\sin(u+z^{(j)}_k+\eta)\sin(u-z^{(j)}_k+\eta)}\no\\[4pt]
&&\qquad\qquad\qquad\times
\lt.\L^{(j+1)}(u+\frac{\eta}{2};\xi^{(j)}-
\!\frac{\eta}{2},\{v^{(j+1)}_k\})\rt\},\no\\
&&\qquad\qquad\qquad\qquad\qquad\qquad\qquad\qquad
\qquad j=1,\ldots,n-2,
\label{Eigenvalue2}\\[6pt]
&&\L^{(n-1)}(u;\xi^{(n-1)})=\tilde{k}_n^{(n-1)}(u)\,
k_n(u;\xi^{(n-1)}).\label{Eigenvalue3}
\end{eqnarray}
The reduced boundary parameters $\{\xi^{(j)}\}$ and inhomogeneous
parameters $\{z^{(j)}_k\}$ are given by
\begin{eqnarray}
 \xi^{(j+1)}=\xi^{(j)}-\frac{\eta}{2},
 \qquad z^{(j+1)}_k=v^{(j)}_k+\frac{\eta}{2},
 \qquad j=0,\ldots,n-2.\label{Parameters}
\end{eqnarray}
Here we have adopted the convention: $\xi=\xi^{(0)}$,
$z^{(0)}_k=z_k$. The complex parameters  $\{v^{(j)}_k\}$ satisfy
the following Bethe ansatz equations:
\begin{eqnarray}
 &&\a^{(1)}(v_s)k_1(v_s;\xi)\frac{\sin(2v_s+\eta)e^{-i\eta}}
 {\sin(2v_s+2\eta)}\no\\[4pt]
 &&\qquad\qquad\qquad\qquad\times \prod_{k\ne s,k=1}^{N_1}
 \frac{\sin(v_s+v_k)\sin(v_s-v_k-\eta)} {\sin(v_s+v_k+2\eta)
 \sin(v_s-v_k+\eta)}\no\\[4pt]
 &&\qquad = \prod_{k=1}^{N} \frac{\sin(v_s+z_k)\sin(v_s-z_k)}
 {\sin(v_s+z_k+\eta)\sin(v_s-z_k+\eta)}\no\\[4pt]
 &&\qquad\qquad\qquad\qquad
 \times\L^{(1)}(v_s+\frac{\eta}{2};\xi-\!\frac{\eta}{2},\{v^{(1)}_k\}),
 \label{BA1}\\[6pt]
 &&\a^{(j+1)}(v^{(j)}_s)k_{j+1}(v^{(j)}_s;\xi^{(j)})\,
 \frac{\sin(2v^{(j)}_s+\eta)e^{-i\eta}}
 {\sin(2v^{(j)}_s+2\eta)}\no\\[4pt]
&&\qquad\qquad\qquad\qquad\times \prod_{k\ne s,k=1}^{N_{j+1}}
 \frac{\sin(v^{(j)}_s+v^{(j)}_k)\sin(v^{(j)}_s-v^{(j)}_k-\eta)}
 {\sin(v^{(j)}_s+v^{(j)}_k+2\eta)\sin(v^{(j)}_s-v^{(j)}_k+\eta)}\no\\[4pt]
 &&\qquad = \prod_{k=1}^{N_j}
 \frac{\sin(v^{(j)}_s+z^{(j)}_k)\sin(v^{(j)}_s-z^{(j)}_k)}
 {\sin(v^{(j)}_s+z^{(j)}_k+\eta)\sin(v^{(j)}_s-z^{(j)}_k+\eta)}\no\\[4pt]
 &&\qquad\qquad\qquad\qquad
 \times\L^{(j+1)}(v^{(j)}_s+\frac{\eta}{2};
 \xi^{(j)}-\!\frac{\eta}{2},\{v^{(j+1)}_k\})
 ,\no\\
 &&\qquad\qquad\qquad\qquad\qquad\qquad\qquad\qquad\qquad  j=1,\ldots,n-2.
 \label{BA2}
\end{eqnarray}

\subsection{Eigenstates and the corresponding eigenvalues}
The relation (\ref{Eq-1}) between $\{H_j\}$ and $\{\t(z_j)\}$ and
the fact that the first term on the r.h.s. of (\ref{trans-2}) is a
c-number enable us to extract the eigenstates  of the generalized
Gaudin operators $\{H_j\}$ and the corresponding eigenvalues from
the results obtained in the previous subsection.

Introduce the functions $\{\b^{(j)}(u,\eta)\}$
\begin{eqnarray}
 \b^{(j+1)}(u,\eta)\equiv
\b^{(j+1)}(u)=\a^{(j+1)}(u)k_{j+1}(u;\xi-\frac{j}{2}\eta),~~j=0,\cdots,
 n-2.\label{function-b}
\end{eqnarray}
Then by (\ref{Eigenvalue2})-(\ref{Parameters}), the Bethe ansatz
equations (\ref{BA1}) and (\ref{BA2}) become, respectively,
\begin{eqnarray}
 &&\hspace{-24pt} \b^{(j+1)}(v^{(j)}_s)\frac{\sin(2v^{(j)}_s+\eta)e^{-i\eta}}{\sin(2v^{(j)}_s+2\eta)}
 \prod_{k\neq s,k=1}^{N_{j+1}}\frac{\sin(v^{(j)}_s+v^{(j)}_k)
 \sin(v^{(j)}_s-v^{(j)}_k-\eta)}{\sin(v^{(j)}_s+v^{(j)}_k+2\eta)
 \sin(v^{(j)}_s-v^{(j)}_k+\eta)}\no\\
 && ~~~~=\b^{(j+2)}(v^{(j)}_s+\frac{\eta}{2})
 \prod_{k=1}^{N_{j}}\frac{\sin(v^{(j)}_s+v^{(j-1)}_k+\frac{\eta}{2})
 \sin(v^{(j)}_s-v^{(j-1)}_k-\frac{\eta}{2})}{\sin(v^{(j)}_s+v^{(j-1)}_k+\frac{3\eta}{2})
 \sin(v^{(j)}_s-v^{(j-1)}_k+\frac{\eta}{2})}\no\\
 &&\qquad\qquad\qquad\qquad\
 \times\prod_{k=1}^{N_{j+2}}\frac{\sin(v^{(j)}_s+v^{(j+1)}_k+\frac{\eta}{2})
 \sin(v^{(j)}_s-v^{(j+1)}_k-\frac{\eta}{2})}{\sin(v^{(j)}_s+v^{(j+1)}_k+\frac{3\eta}{2})
 \sin(v^{(j)}_s-v^{(j+1)}_k+\frac{\eta}{2})},\no\\
&&\hspace{22mm}\qquad\qquad\qquad\qquad\qquad\qquad\qquad\qquad
j=0,\cdots,n-3,\label{BA1-1}\\
 &&
 \hspace{-24pt}
 \b^{(n-1)}(v^{(n-2)}_s)\frac{\sin(2v^{(n-2)}_s+\eta)e^{-i\eta}}{\sin(2v^{(n-2)}_s+2\eta)}
 \prod_{k\neq s,k=1}^{N_{n-1}}\frac{\sin(v^{(n-2)}_s+v^{(n-2)}_k)
 \sin(v^{(n-2)}_s-v^{(n-2)}_k-\eta)}{\sin(v^{(n-2)}_s+v^{(n-2)}_k+2\eta)
 \sin(v^{(n-2)}_s-v^{(n-2)}_k+\eta)}\no\\
 &&~~~~=\tilde{k}_n^{(n-1)}(v_s^{(n-2)}+\frac{\eta}{2})\,\,
 k_n(v_s^{(n-2)}+\frac{\eta}{2};\xi^{(n-1)})\no\\
&&\qquad\qquad\qquad\qquad
\times\prod_{k=1}^{N_{n-2}}\frac{\sin(v^{(n-2)}_s+v^{(n-3)}_k+\frac{\eta}{2})
\sin(v^{(n-2)}_s-v^{(n-3)}_k-\frac{\eta}{2})}{\sin(v^{(n-2)}_s+v^{(n-3)}_k+\frac{3\eta}{2})
 \sin(v^{(n-2)}_s-v^{(n-3)}_k+\frac{\eta}{2})}.\label{BA1-2}
\end{eqnarray}
Here we have used the convention: $v^{(-1)}_k=z_k,\,
k=1,\cdots,N$. The quasi-classical property (\ref{W-1}) of the
R-matrix $W(u)$, (\ref{function-a-1}) and (\ref{function-b}) lead
to the following relations
\begin{eqnarray}
 \b^{(j+1)}(u,0)=1,\quad \frac{\partial}{\partial
 u}\b^{(j+1)}(u,0)=0,\quad j=0,\cdots,n-2.\label{functionb-1}
\end{eqnarray}
Then, one may  introduce functions $\{\g^{(j+1)}(u)\}$ associated
with $\{\b^{(j+1)}(u,\eta)\}$
\begin{eqnarray}
 \g^{(j+1)}(u)=\lt.\frac{\partial}{\partial
 \eta}\b^{(j+1)}(u,\eta)\rt|_{\eta=0},\qquad\qquad
j=0,\cdots,n-2.\label{fuctiong}
\end{eqnarray}
Using (\ref{functionb-1}), we can extract the eigenvalues $h_j$
(resp. the corresponding Bethe ansatz equations) of the Gaudin
operators $H_j$ (\ref{Ham}) from  the expansion around $\eta=0$
for the first order of $\eta$ of the eigenvalues
(\ref{Eigenvalue1}) of the transfer matrix $\t(u=z_j)$ (resp. the
Bethe ansatz equations (\ref{BA1-1}) and (\ref{BA1-2}) ). Finally,
the eigenvalues of the generalized Gaudin operators are
\begin{eqnarray}
h_j=\g^{(1)}(z_j)-\sum_{k=1}^{N_1}
\lt\{\cot(z_j+x_k)+\cot(z_j-x_k)\rt\}.\label{Eig-1}\end{eqnarray}
The parameters
$\{x^{(j)}_k|~k=1,2,\cdots,N_{j+1},~j=0,1,\cdots,n-2\}$ (including
$x_k$ as $x_k=x^{(0)}_k,\,k=1,\cdots, N_1$) are determined by the
following  Bethe ansatz equations:
\begin{eqnarray}
 &&\g^{(j+1)}(x^{(j)}_s)-\cot(2x^{(j)}_s)-i-2\sum_{k\neq
 s,k=1}^{N_{j+1}}\lt\{\cot(x^{(j)}_s+x^{(j)}_k)+
 \cot(x^{(j)}_s-x^{(j)}_k)\rt\}\no\\
 &&\qquad\qquad\quad =\g^{(j+2)}(x^{(j)}_k)-\sum_{k=1}^{N_{j}}
 \lt\{\cot(x^{(j)}_s+x^{(j-1)}_k)+
 \cot(x^{(j)}_s-x^{(j-1)}_k)\rt\}\no\\
 &&\qquad\qquad\qquad\qquad\qquad\ \  -\sum_{k=1}^{N_{j+2}}
 \lt\{\cot(x^{(j)}_s+x^{(j+1)}_k)+
 \cot(x^{(j)}_s-x^{(j+1)}_k)\rt\},\no\\
&&\hspace{27mm}\qquad\qquad\qquad\qquad\qquad\qquad\qquad\qquad
j=0,\cdots,n-3,\label{BAE-1}\\
 &&\g^{(n-1)}(x^{(n-2)}_s)-\cot(2x^{(n-2)}_s)-i-2\sum_{k\neq
 s,k=1}^{N_{n-1}}\lt\{\cot(x^{(n-2)}_s+x^{(n-2)}_k)+
 \cot(x^{(n-2)}_s-x^{(n-2)}_k)\rt\}\no\\
 &&\qquad\qquad\quad =g(x^{(n-2)}_s) -\sum_{k=1}^{N_{n-2}}
 \lt\{\cot(x^{(n-2)}_s+x^{(n-3)}_k)+
 \cot(x^{(n-2)}_s-x^{(n-3)}_k)\rt\}.\label{BAE-2}
\end{eqnarray}
Here we have used the convention: $x^{(-1)}_k=z_k,\,k=1,\cdots,N$
in (\ref{BAE-1}), and the function $g(u)$ is given by  \bea
g(u)=\lt.\frac{\partial}{\partial\eta}\lt(
\tilde{k}_n^{(n-1)}(u+\frac{\eta}{2})\,\,
 k_n(u+\frac{\eta}{2};\xi-\frac{n-1}{2}\eta)\rt)\rt|_{\eta=0}. \eea

%%%%%%%%%%%%%%%%%%%%%%%%%%%%%%%%%%%%%%%%%%%%%%%%%%%%%%%%%%%%%%%%
%                                                              %
%                  Conclusions                                 %
%                                                              %
%%%%%%%%%%%%%%%%%%%%%%%%%%%%%%%%%%%%%%%%%%%%%%%%%%%%%%%%%%%%%%%%

\section{Conclusions} \label{Con}
\setcounter{equation}{0}

We have studied the $A_{n-1}$ Gaudin model with  boundaries
specified by the non-diagonal K-matrices $K^{\pm}(u)$,
(\ref{K-matrix}) and (\ref{DK-matrix}). In addition to the
inhomogeneous parameters $\{z_j\}$, the  Gaudin operators
$\{H_j\}$, (\ref{Ham}), have one discrete parameter $l(1\leq l\leq
n)$ and $n+1$ continuous free parameters
$\{\l_j|j=1,\ldots,n-1;\,\rho,\,\xi\}$ for $1\leq l\leq n-1$ (or
$n$ continuous free parameters $\{\l_j|j=1,\ldots,n-1;\,\rho\}$
for $l=n$). As seen from section 4, although the ``vertex type"
K-matrices $K^{\pm}(u)$ (\ref{K-matrix}) and (\ref{DK-matrix}) are
{\it non-diagonal\/}, the compositions, (\ref{K-F-1}) and
(\ref{K-F-2}), lead to the {\it diagonal\/} ``face-type"
K-matrices after the face-vertex transformation. This has enabled
us to successfully construct the corresponding pseudo-vacuum state
$|\O\rangle$ (\ref{Vac}), diagonalize the generalized Gaudin operators
$\{H_j\}$ by means of  the algebraic Bethe ansatz method, and
derive the eigenvalues (\ref{Eig-1}) as well as the Bethe ansatz
equations (\ref{BAE-1}) and (\ref{BAE-2}) of the boundary Gaudin model.

%%%%%%%%%%%%%%%%%%%%%%%%%%%%%%%%%%%%%%%%%%%%%%%%%%%%%%%%%%%%%%%
%                                                             %
%  Acknowledgments                                            %
%                                                             %
%%%%%%%%%%%%%%%%%%%%%%%%%%%%%%%%%%%%%%%%%%%%%%%%%%%%%%%%%%%%%%%
\section*{Acknowledgements}
The financial support from  Australian Research Council through
the Discovery-Projects and Linkage-International grants  is
gratefully acknowledged. The first two authors  would like to
thank Yukawa Institute for Theoretical Physics, Kyoto University
for kind hospitality during their visit to Kyoto.

%%%%%%%%%%%%%%%%%%%%%%%%%%%%%%%%%%%%%%%%%%%%%%%%%%%%%%%%%%%%%
%                                                           %
%                Appendix                                   %
%                                                           %
%%%%%%%%%%%%%%%%%%%%%%%%%%%%%%%%%%%%%%%%%%%%%%%%%%%%%%%%%%%%%

\section*{Appendix: Explicit matrix forms of K-matrices}
\setcounter{equation}{0}
\renewcommand{\theequation}{A.\arabic{equation}}

In this appendix, we  give the explicit matrix expressions of the
K-matrices (\ref{K-matrix}) for the cases $n=3,4$.

\vskip0.2in
 \noindent {\large \bf The $A^{(1)}_2$ case:}
\vskip0.2in

 \noindent There are three types of K-matrices $K^-(u)$, which are labelled
 by the discrete parameters $l$, for the
 trigonometric $A^{(1)}_{2}$ model.
\begin{itemize}
\item For the case of $l=1$, the $7$ non-vanishing matrix elements
$K^-(u)^k_j$ are given by: \bea
&&K^-(u)^1_1=\frac{e^{2iu}}{e^{2iu}+e^{\rho}}
\lt(1-e^{\rho}\frac{\sin(u-\xi)}{\sin(u+\xi)}\rt),\no\\
&&K^-(u)^1_2=\frac{e^{-2i\l_1+\rho}}{e^{2iu}+e^{\rho}}
\lt(1+e^{2iu}\frac{\sin(u-\xi)}{\sin(u+\xi)}\rt),\no\\
&&K^-(u)^1_3=-\frac{e^{-2i(\l_1+\l_2)+\rho}}{e^{2iu}+e^{\rho}}
\lt(1+e^{2iu}\frac{\sin(u-\xi)}{\sin(u+\xi)}\rt),\no\\
&&K^-(u)^2_1=\frac{e^{2i\l_1+i(u+\xi)}\sin2u}
{\lt(e^{2iu}+e^{\rho}\rt)\sin(u+\xi)},\no\\
&&K^-(u)^2_2=\frac{1}{e^{2iu}+e^{\rho}}
\lt(e^{\rho}-\frac{\sin(u-\xi)}{\sin(u+\xi)}\rt),\no\\
&&K^-(u)^2_3=-\frac{e^{-2i\l_2-i(u-\xi)+\rho}\sin2u}{\lt(e^{2iu}+e^{\rho}\rt)
\sin(u+\xi)},\,\,
K^-(u)^3_3=e^{-2iu}\frac{\sin(\xi-u)}{\sin(\xi+u)}.\label{K1-1}
\eea

\item For the case of $l=2$, the $7$ non-vanishing matrix elements
$K^-(u)^k_j$ are given by: \bea
%&&K^-(u)^1_1=\frac{e^{2iu}}{e^{2iu}+e^{\rho}}
%\lt(1-e^{\rho}\frac{\sin(u-\xi)}{\sin(u+\xi)}\rt),\no\\
%&&K^-(u)^1_2=\frac{e^{-2i\l_1+\rho}}{e^{2iu}+e^{\rho}}
%\lt(1+e^{2iu}\frac{\sin(u-\xi)}{\sin(u+\xi)}\rt),\no\\
%&&K^-(u)^1_3=-\frac{e^{-2i(\l_1+\l_2)+\rho}}{e^{2iu}+e^{\rho}}
%\lt(1+e^{2iu}\frac{\sin(u-\xi)}{\sin(u+\xi)}\rt),\,\,\,K^-(u)^2_2=1,\no\\
&&K^-(u)^1_1,\,K^-(u)^1_2,\,K^-(u)^1_3\,\,{\rm
are\,the\,same\,as\,those\,in}\,(\ref{K1-1}),\no\\
&&K^-(u)^2_2=1,\,\,K^-(u)^3_1=-\frac{e^{2i(\l_1+\l_2)+i(u+\xi)}\sin2u}
{\lt(e^{2iu}+e^{\rho}\rt)\sin(u+\xi)},\no\\
&&K^-(u)^3_2=\frac{e^{2i\l_2+i(u+\xi)}\sin2u}{\lt(e^{2iu}+e^{\rho}\rt)
\sin(u+\xi)},\no\\
&&K^-(u)^3_3=\frac{1}{e^{2iu}+e^{\rho}}
\lt(e^{\rho}-\frac{\sin(u-\xi)}{\sin(u+\xi)}\rt). \eea

\item For the case of $l=3$, the $5$ non-vanishing matrix elements
$K^-(u)^k_j$ are given by: \bea
&&K^-(u)^1_1=\frac{e^{2iu}+e^{4iu+\rho}}{e^{2iu}+e^{\rho}},\,\,
K^-(u)^1_2=-\frac{e^{-2i\l_1+\rho}\lt(e^{4iu}-1\rt)}{e^{2iu}+e^{\rho}},\no\\
&&K^-(u)^1_3=\frac{e^{-2i(\l_1+\l_2)+\rho}\lt(e^{4iu}-1\rt)}{e^{2iu}+e^{\rho}},
\,\,K^-(u)^2_2=K^-(u)^3_3=1.\label{K1-2}\eea

\end{itemize}

\noindent {\large\bf The $A^{(1)}_3$ case:}

\vskip0.2in

\noindent There are four types of K-matrices $K^-(u)$, which are
labelled by the discrete parameters $l$, for the trigonometric
$A^{(1)}_{3}$ model.
\begin{itemize}
\item For the case of $l=1$, the $10$ non-vanishing matrix
elements $K^-(u)^k_j$  are given by:
\bea &&K^-(u)^1_1=\frac{e^{2iu}}{e^{2iu}-e^{\rho}}
\lt(1+e^{\rho}\frac{\sin(u-\xi)}{\sin(u+\xi)}\rt),\no\\
&&K^-(u)^1_2=-\frac{e^{-2i\l_1+\rho}}{e^{2iu}-e^{\rho}}
\lt(1+e^{2iu}\frac{\sin(u-\xi)}{\sin(u+\xi)}\rt),\no\\
&&K^-(u)^1_3=\frac{e^{-2i(\l_1+\l_2)+\rho}}{e^{2iu}-e^{\rho}}
\lt(1+e^{2iu}\frac{\sin(u-\xi)}{\sin(u+\xi)}\rt),\,\,\no\\
&&K^-(u)^1_4=-\frac{e^{-2i(\l_1+\l_2+\l_3)+\rho}}{e^{2iu}-e^{\rho}}
\lt(1+e^{2iu}\frac{\sin(u-\xi)}{\sin(u+\xi)}\rt),\no\\
&&K^-(u)^2_1=\frac{e^{2i\l_1+i(u+\xi)}\sin2u}
{\lt(e^{2iu}-e^{\rho}\rt)\sin(u+\xi)},\no\\
&&K^-(u)^2_2=-\frac{1}{e^{2iu}-e^{\rho}}
\lt(e^{\rho}+\frac{\sin(u-\xi)}{\sin(u+\xi)}\rt),\no\\
&&K^-(u)^2_3=\frac{e^{-2i\l_2-i(u-\xi)+\rho}\sin2u}{\lt(e^{2iu}-e^{\rho}\rt)
\sin(u+\xi)},\no\\
&&K^-(u)^2_4=-\frac{e^{-2i(\l_2+\l_3)-i(u-\xi)+\rho}\sin2u}{\lt(e^{2iu}-e^{\rho}\rt)
\sin(u+\xi)},\no\\
&&K^-(u)^3_3=K^-(u)^4_4=e^{-2iu}\frac{\sin(\xi-u)}{\sin(\xi+u)}.
\label{K2-1}\eea

\item For the case of $l=2$, the $10$ non-vanishing matrix
elements $K^-(u)^k_j$  are given by:
\bea
%&&K^-(u)^1_1=\frac{e^{2iu}}{e^{2iu}-e^{\rho}}
%\lt(1+e^{\rho}\frac{\sin(u-\xi)}{\sin(u+\xi)}\rt),\no\\
%&&K^-(u)^1_2=-\frac{e^{-2i\l_1+\rho}}{e^{2iu}-e^{\rho}}
%\lt(1+e^{2iu}\frac{\sin(u-\xi)}{\sin(u+\xi)}\rt),\no\\
%&&K^-(u)^1_3=\frac{e^{-2i(\l_1+\l_2)+\rho}}{e^{2iu}-e^{\rho}}
%\lt(1+e^{2iu}\frac{\sin(u-\xi)}{\sin(u+\xi)}\rt),\,\,\no\\
%&&K^-(u)^1_4=-\frac{e^{-2i(\l_1+\l_2+\l_3)+\rho}}{e^{2iu}-e^{\rho}}
%\lt(1+e^{2iu}\frac{\sin(u-\xi)}{\sin(u+\xi)}\rt),\,\,K^-(u)^2_2=1,\no\\
&&K^-(u)^1_1,\,K^-(u)^1_2,\,K^-(u)^1_3,\,K^-(u)^1_4\,\,{\rm
are\,
the\,same\,as\,those\,in\,} (\ref{K2-1})\no\\
&&K^-(u)^2_2=1,\,\,K^-(u)^3_1=-\frac{e^{2i(\l_1+\l_2)+i(u+\xi)}\sin2u}
{\lt(e^{2iu}-e^{\rho}\rt)\sin(u+\xi)},\no\\
&&K^-(u)^3_2=\frac{e^{2i(\l_2)+i(u+\xi)}\sin2u}
{\lt(e^{2iu}-e^{\rho}\rt)\sin(u+\xi)},\no\\
&&K^-(u)^3_3=-\frac{1}{e^{2iu}-e^{\rho}}
\lt(e^{\rho}+\frac{\sin(u-\xi)}{\sin(u+\xi)}\rt),\no\\
&&K^-(u)^3_4=\frac{e^{-2i\l_3-i(u-\xi)+\rho}\sin2u}{\lt(e^{2iu}-e^{\rho}\rt)
\sin(u+\xi)},\,\,K^-(u)^4_4=e^{-2iu}\frac{\sin(\xi-u)}{\sin(\xi+u)}.
\eea

\item For the case of $l=3$, the $10$ non-vanishing matrix
elements $K^-(u)^k_j$  are given by:
\bea
%&&K^-(u)^1_1=\frac{e^{2iu}}{e^{2iu}-e^{\rho}}
%\lt(1+e^{\rho}\frac{\sin(u-\xi)}{\sin(u+\xi)}\rt),\no\\
%&&K^-(u)^1_2=-\frac{e^{-2i\l_1+\rho}}{e^{2iu}-e^{\rho}}
%\lt(1+e^{2iu}\frac{\sin(u-\xi)}{\sin(u+\xi)}\rt),\no\\
%&&K^-(u)^1_3=\frac{e^{-2i(\l_1+\l_2)+\rho}}{e^{2iu}-e^{\rho}}
%\lt(1+e^{2iu}\frac{\sin(u-\xi)}{\sin(u+\xi)}\rt),\,\,\no\\
%&&K^-(u)^1_4=-\frac{e^{-2i(\l_1+\l_2+\l_3)+\rho}}{e^{2iu}-e^{\rho}}
%\lt(1+e^{2iu}\frac{\sin(u-\xi)}{\sin(u+\xi)}\rt),\no\\
&&K^-(u)^1_1,\,K^-(u)^1_2,\,K^-(u)^1_3,\,K^-(u)^1_4\,\,{\rm are\,
the\,same\,as\,those\,in\,} (\ref{K2-1})\no\\
&&K^-(u)^2_2=K(u)^3_3=1,\no\\
&&K^-(u)^4_1=\frac{e^{2i(\l_1+\l_2+\l_3)+i(u+\xi)}\sin2u}
{\lt(e^{2iu}-e^{\rho}\rt)\sin(u+\xi)},\no\\
&&K^-(u)^4_2=-\frac{e^{2i(\l_2+\l_3)+i(u+\xi)}\sin2u}
{\lt(e^{2iu}-e^{\rho}\rt)\sin(u+\xi)},\no\\
&&K^-(u)^4_3=\frac{e^{2i\l_3+i(u+\xi)}\sin2u}{\lt(e^{2iu}-e^{\rho}\rt)
\sin(u+\xi)},\no\\
&&K^-(u)^4_4=-\frac{1}{e^{2iu}-e^{\rho}}
\lt(e^{\rho}+\frac{\sin(u-\xi)}{\sin(u+\xi)}\rt).\eea

\item For the case of $l=4$, the $7$ non-vanishing matrix elements
$K^-(u)^k_j$  are given by: \bea
&&K^-(u)^1_1=\frac{e^{2iu}-e^{4iu+\rho}}{e^{2iu}-e^{\rho}},\,\,
K^-(u)^1_2=\frac{e^{-2i\l_1+\rho}\lt(e^{4iu}-1\rt)}{e^{2iu}-e^{\rho}},\no\\
&&K^-(u)^1_3=-\frac{e^{-2i(\l_1+\l_2)+\rho}\lt(e^{4iu}-1\rt)}{e^{2iu}-e^{\rho}},\no\\
&&K^-(u)^1_4=\frac{e^{-2i(\l_1+\l_2+\l_3)+\rho}\lt(e^{4iu}-1\rt)}
{e^{2iu}-e^{\rho}},\no\\
&&K^-(u)^2_2=K^-(u)^3_3=K^-(u)^4_4=1.\label{K2-2}\eea

\end{itemize}
The above explicit results confirm the following properties for
the non-diagonal K-matrices of the $A^{(1)}_{n-1}$ vertex model:
for $l=1,\ldots,n-1$, $K^-(u)$ depends on $n+1$ continuous free
parameters and has $3n-2$ non-vanishing matrix elements; for
$l=n$, $K^-(u)$ depends on $n$ continuous free parameters and has
$2n-1$ non-vanishing matrix elements. So our K-matrices contain
more boundary parameters and more non-vanishing matrix elements
than those found in \cite{Aba95,Lim02,Lim04}.

%%%%%%%%%%%%%%%%%%%%%%%%%%%%%%%%%%%%%%%%%%%%%%%%%%%%%%%%%%%%%%%
%                                                             %
%  References                                                 %
%                                                             %
%%%%%%%%%%%%%%%%%%%%%%%%%%%%%%%%%%%%%%%%%%%%%%%%%%%%%%%%%%%%%%%

\end{document}